\documentclass[12pt,titlepage]{article}
\usepackage[letterpaper,top=1in,bottom=1in,left=1.25in,right=1.25in]{geometry}
\usepackage{graphicx,cite,url}

\usepackage{color}
\usepackage{hyperref}
\hypersetup{
    colorlinks=true,
    linkcolor=blue,
    filecolor=magenta,
    urlcolor=blue,
}

\usepackage{graphicx}
\usepackage{subfigure}
\usepackage{bm}
\usepackage{amssymb}
\usepackage{cite}
\usepackage{stfloats}
\usepackage{bm}
\usepackage{mathrsfs,amsmath,cleveref}
\usepackage{subfigure}

\DeclareGraphicsExtensions{.eps,.ps,.pdf,.bmp}

\begin{document}

\title{High-speed $\emph{in vitro}$ intensity diffraction tomography}

\author{Jiaji Li$^{1,\dagger}$, Alex Matlock$^{2,\dagger}$, Yunzhe Li$^{2}$, Qian Chen$^{1}$, Chao Zuo$^{1,*}$, Lei Tian$^{2,*}$
\\
\multicolumn{1}{p{\textwidth}}{\centering\emph{\normalsize
1. School of Electronic and Optical Engineering, Nanjing University of Science and Technology, No. 200 Xiaolingwei Street, Nanjing, Jiangsu Province 210094, China\\
2. Department of Electrical and Computer Engineering, Boston University, Boston, MA 02215, USA\\
$\dagger$, Equal contribution\\
}}}

\maketitle
\begin{abstract}
We demonstrate a label-free, scan-free {\it intensity} diffraction tomography technique utilizing annular illumination (aIDT) to rapidly characterize large-volume 3D refractive index distributions \emph{in vitro}.
By optimally matching the illumination geometry to the microscope pupil, our technique reduces the data requirement by 60$\times$ to achieve high-speed 10 Hz volume rates.
Using 8 intensity images, we recover \mbox{$\sim350\times100\times20\mu$m$^3$} volumes with near diffraction-limited lateral resolution of 487 nm and axial resolution of 3.4 $\mu$m.
Our technique's large volume rate and high resolution enables 3D quantitative phase imaging of complex living biological samples across multiple length scales.
We demonstrate aIDT's capabilities on unicellular diatom microalgae, epithelial buccal cell clusters with native bacteria, and live \emph{Caenorhabditis elegans} specimens.
Within these samples, we recover macro-scale cellular structures, subcellular organelles, and dynamic micro-organism tissues with minimal motion artifacts.
Quantifying such features has significant utility in oncology, immunology, and cellular pathophysiology, where these morphological features are evaluated for changes in the presence of disease, parasites, and new drug treatments.
Finally, we simulate our aIDT system to highlight the accuracy and sensitivity of our technique.
aIDT shows promise as a powerful high-speed, label-free computational microscopy technique applications where natural imaging is required to evaluate environmental effects on a sample in real-time.
We provide example datasets and an open source implementation of aIDT at \href{https://github.com/bu-cisl/IDT-using-Annular-Illumination}{https://github.com/bu-cisl/IDT-using-Annular-Illumination}.
\end{abstract}

\section{Introduction}
Three-dimensional (3D) refractive index (RI) distributions of cells and tissues are useful for the morphological detection and diagnosis of disease in biomedical imaging \cite{park2018quantitative}.
For example, characterizing RI distribution has shown particular utility in understanding disease and morphogenesis \cite{merola2017tomographic}.
Quantitatively characterizing thick biological samples across multiple subcellular and multicellular scales, however, remains a challenging task.
Here, we present a scan-free, high-speed intensity diffraction tomography (IDT) technique based on a standard microscope modified with an annular LED illumination hardware unit.
Our label-free imaging method enables \emph{in vitro} biological sample observation providing intrinsic 3D structural RI sensitivity of large volumes at real-time acquisition speeds.

Due to the low absorption and contrast of biological samples in the visible spectrum, exogenous labels (e.g. fluorophores) are commonly used as biomarkers to visualize regions of interest.
For example, confocal fluorescence and two-photon microscopy are commonly used when imaging thick 3D samples.
Although these methods provide excellent optical sectioning, the excitation light
and contrast agents for fluorescence imaging can cause photobleaching, phototoxicity, and other damaging effects that artificially alter the sample's behavior and structure \cite{Stephens2003}.
These factors have pushed the need for label-free microscopy, where biological samples are studied in their natural states.
Quantitative phase imaging (QPI) \cite{park2018quantitative,popescu2011quantitative} is one such technique that measures an incident field's phase shifts induced by the sample to recover the sample's physical properties without staining or labeling.
Both interferometry~\cite{kemper2008digital,ferraro2006quantitative,Wang:11,baek2019kramers} and non-interferometry based \cite{Waller:2010aa,Alieva:2014hb,Tian2015a, zuo2017high,li2017efficient} QPI techniques have been developed to recover a sample's phase map in two-dimensions (2D).
A single 2D integrated phase image, however, is insufficient for characterizing 3D heterogeneous samples.
Recently developed 3D phase tomography techniques instead recover the sample's 3D RI distribution to visualize intracellular structures \cite{Choi.etal2007,cotte2013marker,nguyen2017gradient}.

The most widely used interferometry-based RI tomography technique is optical diffraction tomography (ODT).
In ODT, the scattered field is directly measured from digitally recorded interferograms taken under different illumination angles (i.e. phase projection measurement).
Several ODT approaches have been applied in biomedical studies for evaluating cell physiology~\cite{Sung.etal2009,kus2014tomographic}, pathophysiology and immunology~\cite{chandramohanadas2011biophysics,yoon2015label,lee2013quantitative,yakimovich2018label},
oncology~\cite{sung2012stain,li2019quantitative,zhikhoreva2018morphological}, and micro-organism and intracellular particle tracking~\cite{Isikman2011,kim2018label}.
Notably, annular illumination is particularly effective in achieving high-quality 3D RI recovery using a relatively small number of phase projection measurements.
The phase nanoscopy technique~\cite{cotte2013marker} allows both static and time-lapse 3D RI recovery of cells by mechanically scanning a laser illumination unit along a ring-shaped trajectory.
An alternative ODT microscope builds on a programmable illumination unit using a digital micromirror device (DMD) to generate plane wave illumination with annularly distributed angles \cite{shin2015active,shin2018super}.
However, both techniques require additional optical paths for off-axis or common-path interferometric recording that are difficult to implement in existing microscope platforms.

The alternative approach we demonstrate here uses {\it intensity-only, non-interferometry} based measurements for quantitative phase imaging based on the principle of IDT~\cite{ling2018high}.
Other intensity-based approaches have been explored but often require sample or objective scanning along the axial direction to capture a through-focus intensity stack \cite{Soto2017,Rodrigo2017,li2018three}.
The 3D RI map is then recovered from this image stack using deconvolution algorithms \cite{Jenkins2015,Chen2016}.
For example, the gradient light interference microscopy~\cite{nguyen2017gradient} achieves 3D imaging of thick biological samples using a differential interference contrast (DIC) through-focus intensity stack.
The 3D differential phase contrast microscopy approach takes an axially-scanned intensity stack under asymmetric partially coherent illumination~\cite{Chen2016}.
In these cases, however, the mechanical motion reduces the measurement's temporal resolution.
Recently, {\it motion-free} IDT techniques have been demonstrated by our group allowing 3D RI information recovery using oblique illumination-only intensity measurements {\it without} any mechanical scanning.
While lasers coupled with DMDs, galvanometers, and other programmable sources have been used for scan-free oblique illumination~\cite{cotte2013marker, shin2015active, Eckert_2018}, we equip a standard brightfield microscope with a ring LED array.
The LED array provides a low-cost illumination source easily adoptable for biological research that avoids the speckle artifacts present with coherent lasers.
Unfortunately, our original illumination-based IDT technique~\cite{ling2018high} still exhibited poor temporal resolution due to its large illumination quantity requiring $>600$ intensity images over several minutes per measurement acquisition.

To improve IDT's temporal resolution, we develop a fast and accurate annular illumination IDT (aIDT) technique overcoming these limitations.
Importantly, our technique reduces the data requirement by more than 60$\times$, achieving more than 10 Hz for imaging a \mbox{$\sim350\times100\times20\mu$m$^3$} volume with  near diffraction-limited lateral resolution of 487 nm and axial resolution of 3.4 $\mu$m in the 3D RI reconstruction.
These improvements enable {\it in vitro} dynamic 3D RI characterizations of living biological samples.
We show that our technique provides useful subcellular information on multiple species including unicellular diatom microalgae, clustered epithelial buccal cells, and live \emph{Caenorhabditis elegans} (\emph{C. elegans}) multi-cellular specimens.

Our technique is enabled by several hardware and algorithmic innovations.
First, our hardware employs a programmable LED ring consisting of only 24 surface-mounted LEDs.
Compared to existing LED matrix based systems \cite{Zheng2013,Tian2015b,Tian.Waller2015,Tian2014a,Chen2016,Tian2014},
the reduced number of LEDs significantly speed up the acquisition.
More importantly, the ring illumination geometry can optimally fit with the circular microscope pupil, which we show is crucial for both data reduction  and high-quality RI reconstruction.
We develop an illumination-based IDT theory that highlights the optimal imaging condition achieved by  matching the illumination angle with the objective numerical aperture (NA).
This illumination scheme optimally encodes both low and high spatial frequency RI information across the entire 3D volume using a small number of intensity measurements.
This allows us to use only 8 intensity measurements for imaging fast dynamics and provides an optimal trade-off between RI reconstruction quality and motion artifacts.
We develop a new reconstruction algorithm that implements a non-iterative slice-wise deconvolution that is both data and computationally efficient.
We further develop a robust self-calibration algorithm to correct for the LED positions, which we show is critical for high-quantity 3D RI reconstruction, in particular at large defocus positions.
Finally, we simulate our aIDT system to show accurate reconstructions within $1\times10^{-3}$ of the object's true RI and sensitivity to RI changes of $2\times10^{-4}$ under our system's SNR for weakly scattering objects.

\section{Methods}
\subsection{aIDT principle}

\begin{figure*}[!t]
    \centering
    \includegraphics[width=1\textwidth]{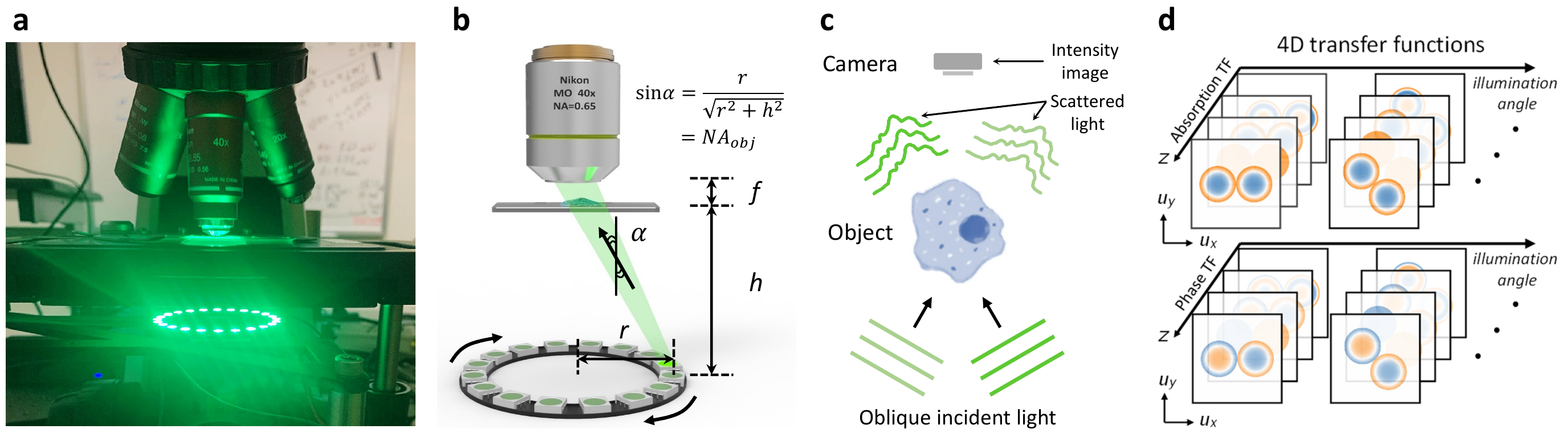}
    \caption{Illustration of our aIDT imaging system.
    (a) A photo of our aIDT system consisting of a standard microscope equipped with an LED ring.
    A visualization demonstrating the operation of the system is shown in Video 1.
    (b) An LED ring illumination unit is placed underneath the sample.  The distance $h$ is tuned such that the illumination angle $\alpha$ is matched with the objective NA.
    (c) Each IDT image measures the interference between the scattered and the unperturbed fields.
    (d) The absorption and phase transfer functions at various illumination angles and sample depths
    (Video 1, MPEG, 2.7 MB).
        }    \label{Fig1}
\end{figure*}

Our aIDT imaging system combines an LED ring illumination unit and a standard brightfield microscope (Fig.~\ref{Fig1}a).
The LED ring is placed some distance $h$ away under the sample, as illustrated in Fig.~\ref{Fig1}b.
Importantly, the distance is carefully tuned such that the ring is matched with the perimeter of the objective lens's pupil aperture.
This can be done because this illumination design approximately follows the K{\"o}hler geometry, in which each LED provides a plane wave of unique angle.
Denoting the radius of the ring as $r$, the illumination NA (NA$_{\text{illum}}$) of each LED is set by NA$_{\text{illum}}$ = ${r \mathord{\left/ {\vphantom {r {\sqrt {{r^2} + {h^2}} }}} \right. \kern-\nulldelimiterspace} {\sqrt {{r^2} + {h^2}} }}$.
Hence, one can achieve NA$_{\text{illum}}$ matching the objective NA (i.e. NA$_{\text{obj}}$ = NA$_{\text{illum}}$) by simply adjusting the LED height $h$.
During the experiment, we acquire up to 24 images capturing brightfield intensity measurements from each individual LED in our ring unit (Fig.~\ref{Fig1}c).
By downsampling the total LED number, we can improve our acquisition speed to accommodate dynamic live samples.

The intensity of each image results from the interference of the scattered field from a weakly-scattering object and the unperturbed illuminating field (Fig.~\ref{Fig1}c).
By quantifying the Fourier space information under the first Born approximation, we derive phase and absorption transfer functions~(TFs)~\cite{ling2018high} {\it linearly} relating the object's complex permittivity distribution to the measured intensity.
The exact form of the TFs are detailed in Eqs.~(\ref{Eq1a}--\ref{Eq1b}) in Sec.~\ref{subsection:datapreprocessing}.
The TFs are angle and depth-dependent and result in the 4D Fourier space representation shown in Fig.~\ref{Fig1}d.
This relation allows us to implement a fast and efficient slice-wise 3D deconvolution algorithm with a closed form solution (Eqs.~(\ref{Eq2a}--\ref{Eq2b}) in Sec.~\ref{subsection:deconvolution}) to recover the complex 3D RI distribution.

Several important observations govern our illumination design.
First, each TF's Fourier coverage features a pair of circular regions describing the scattered field's information and its complex conjugate.
The system's objective NA$_{\text{obj}}$ defines the circles' radius while the illumination angle NA$_{\text{illum}}$ defines the center positions.
By matching  NA$_{\text{illum}}$ to NA$_{\text{obj}}$, one ensures maximizing the Fourier coverage allowed by the system (i.e. 2NA$_{\text{obj}}$), as demonstrated in supplementary Fig.~S1.
Second, the phase information is captured by the {\it anti-symmetric} Fourier information whereas the absorption information is by {\it symmetric} information (Fig.~\ref{Fig1}d).
The anti-symmetry in the phase TF further indicates that any overlap between the two circular regions (by using NA$_{\text{illum}}<$ NA$_{\text{obj}}$) will cancel the captured low spatial frequency information (Fig.~ S1a) ~\cite{ling2018high, Tian2015a}.
Setting NA$_{\text{illum}}=$ NA$_{\text{obj}}$ removes this overlap and ensures optimal low frequency phase information coverage.
Together, our annular illumination scheme captures both high and low spatial frequency phase and absorption information in 3D using a small number of intensity-only measurements.

\subsection{Transfer function analysis}
\label{subsection:datapreprocessing}

In the aIDT forward model, a 3D sample is discretized into a stack of 2D sample slices.
Following the IDT derivation~\cite{ling2018high}, the 3D phase and absorption TFs can be derived slice-by-slice and as a function of the illumination angle.
The analytical expression of the phase TF ($H_P$) and absorption TF ($H_A$) can be expressed as
\begin{subequations}
\begin{align}
{H_P}\left( {\bf{u}} \right) = \frac{{i\Delta zk_0^2}}{2}\left\{ {P\left( {{\bf{u}} - {\rho _{\bf{s}}}} \right)\frac{{\exp \left[ { - i\left( {\eta \left( {{\bf{u}} - {\rho _{\bf{s}}}} \right) - {\eta _{\bf{s}}}} \right)z} \right]}}{{\eta \left( {{\bf{u}} - {\rho _{\bf{s}}}} \right)}} - P\left( {{\bf{u}} + {\rho _{\bf{s}}}} \right)\frac{{\exp \left[ {i\left( {\eta \left( {{\bf{u}} + {\rho _{\bf{s}}}} \right) - {\eta _{\bf{s}}}} \right)z} \right]}}{{\eta \left( {{\bf{u}} + {\rho _{\bf{s}}}} \right)}}} \right\},
   \label{Eq1a}
\\
{H_A}\left( {\bf{u}} \right) =  - \frac{{\Delta zk_0^2}}{2}\left\{ {P\left( {{\bf{u}} - {\rho _{\bf{s}}}} \right)\frac{{\exp \left[ { - i\left( {\eta \left( {{\bf{u}} - {\rho _{\bf{s}}}} \right) - {\eta _{\bf{s}}}} \right)z} \right]}}{{\eta \left( {{\bf{u}} - {\rho _{\bf{s}}}} \right)}} + P\left( {{\bf{u}} + {\rho _{\bf{s}}}} \right)\frac{{\exp \left[ {i\left( {\eta \left( {{\bf{u}} + {\rho _{\bf{s}}}} \right) - {\eta _{\bf{s}}}} \right)z} \right]}}{{\eta \left( {{\bf{u}} + {\rho _{\bf{s}}}} \right)}}} \right\},
 \label{Eq1b}
\end{align}
\end{subequations}
where $\bf{u}$ denotes the 2D spatial frequency variable, $\Delta z$ is the axial sampling spacing (i.e. slice spacing), $k_0=2\pi/\lambda$ the wave number, $\lambda$ the illumination wavelength, $P$ the pupil function of the objective, ${{\bf{\rho }}_{\bf{s}}}$ the spatial frequency of the illumination, $\eta \left( {\bf{u}} \right) = \sqrt {k_0^2 - {{\left| {\bf{u}} \right|}^2}} $ the axial frequency, and $z$ the sample's axial location.

\subsection{3D RI reconstruction algorithm}
\label{subsection:deconvolution}

To achieve 3D RI reconstruction, aIDT solves an inverse problem through deconvolution.
First, each intensity image is processed to remove the background.
Next, the Tikhonov regularized deconvolution is performed to reconstruct the object's real and imaginary RI.
The main idea of our slice-wise deconvolution process is to replace the continuous integration along the axial direction by a discrete sum over the slice index.
Importantly, since the TFs are independent between slices, the scattering information from different sample slices is also {\it decoupled}.
This decoupling allows us to use a computationally efficient, slice-based deconvolution procedure to reconstruct the cross-sectional RI distribution one slice at a time.
The slice spacing is chosen arbitrarily during the computation.
The achieved axial resolution is characterized by analyzing the reconstructed stack and is found to approximately match with the diffraction limit, set by
${\lambda}/({{n_m} - \sqrt {n_m^2 - \mathrm{NA}_{\mathrm{obj}}^2} })$, where $n_m$ is the RI of surrounding medium.
This closed-form solutions for real part of the permittivity contrast (i.e. phase) $\Delta {\varepsilon _{\mathrm{Re}}}$ and imaginary part of the permittivity contrast (i.e. absorption) $\Delta {\varepsilon _{\mathrm{Im}}}$ at each axial slice are

\begin{subequations}
\begin{align}
\label{Eq2a}
\Delta {\varepsilon _{\mathrm{Re}}}[m] = {\mathscr{F}}^{-1}\Bigg\{\frac{1}{A}\Bigg\{
&\bigg( \sum\limits_l{{|H_A(l,m)|}^2}+\beta\bigg) \odot \bigg(\sum\limits_l {H_P^*(l,m) \odot \widetilde{g}[l]}\bigg)- \nonumber\\
& \bigg(\sum\limits_l {{H_P}^*(l,m) \odot H_A(l,m)} \bigg)\odot \bigg(\sum\limits_l {H_A^*(l,m)\odot \widetilde{g}[l]}\bigg)\Bigg\}\Bigg\} \\
\label{Eq2b}
\Delta {\varepsilon _{\mathrm{Im}}}[m] = {\mathscr{F}}^{-1}\Bigg\{\frac{1}{A}\Bigg\{
&\bigg( \sum\limits_l{{|H_P(l,m)|}^2}+\alpha\bigg) \odot \bigg(\sum\limits_l {H_A^*(l,m) \odot \widetilde{g}[l]}\bigg)- \nonumber\\
& \bigg(\sum\limits_l {{H_P}(l,m) \odot H_A^*(l,m)} \bigg)\odot \bigg(\sum\limits_l {H_P^*(l,m)\odot \widetilde{g}[l]}\bigg)\Bigg\}\Bigg\}
\end{align}
\end{subequations}
Where ${\mathscr{F}}^{-1}$ denotes the 2D inverse Fourier transform (IFT); $A$  is a normalization term $ \Big[\sum\limits_l{{|H_P(l,m)|}^2}+\alpha\Big] \odot \Big[ \sum\limits_l{{|H_A(l,m)|}^2}+\beta\Big]-\Big[\sum\limits_l {{H_P}(l,m) \odot H_A^*(l,m)}\Big]\odot\Big[\sum\limits_l {{H_P}^*(l,m) \odot H_A(l,m)} \Big] $; $\left[ m \right]$ indexes the $m$th sample slice, $\left[ l \right]$  the $l$th intensity image, ${H_{\rm{A}}}\left( {l,{\rm{m}}} \right)$  and  ${H_P}\left( {l,{\rm{m}}} \right)$ are the discretized 4D TFs for the $m$th slice from the $l$th illumination; $\odot$ denotes element-wise multiplication between two matrices, $\alpha$ and $\beta$  are the regularization parameters for the phase and absorption, and $\widetilde{g}[l]$ is the Fourier spectrum of the background-subtracted intensity image.

\subsection{System setup}

Our aIDT setup is built on an upright brightfield microscope (E200, Nikon) and replaces the existing illumination unit with a ring LED (1586, Adafruit).
The radius of the ring LED unit is $\sim30$ mm.
The LED ring is placed $\sim35$mm away from the sample, whose center is aligned with the optical axis of the microscope.
Each LED approximately  provides  spatially coherent quasi-monochromatic illumination with central wavelength $\lambda$ = 515 nm, and $\sim20$nm bandwidth.
All experiments in the main text were conducted using a 40$\times$ microscope objective (MO) (0.65 NA, CFI Plan Achro).
Images were taken with an sCMOS camera (PCO. Panda 4.2, 6.5$\mu$m pixel size), which is synchronized with the LED source to provide camera limited acquisition speed.
The LED ring uint is driven by a micro-controller (Arduino Uno).
In addition, we provide additional results using a 10$\times$ (0.25 NA, CFI Plan Achro) MO to image spiral algae (S68786, Fisher Scientific) in supplementary material and Video S1.

During the experiments of imaging living \emph{C. elegans}, a rectangular field of view consisting of 2048$\times$600 pixels is optimized  to match the size of the sample and achieve 85 Hz acquisition speed.
All the processing is done using MATLAB 2018b on a personal computer. The processing time to perform  LED position calibration  (using a 400$\times$400$\times$24 intensity stack) is about 2 seconds. 3D reconstruction time for a 1024$\times$1024$\times$51 RI stack takes about 50 seconds.

\subsection{Self-calibration method}
\label{subsection:selfcal}

During the reconstruction, we first perform a numerical self-calibration procedure.
The assumed illumination angles for the ring LED do not necessarily match the experimental implementation, and the use of incorrect angles can lead to significant reconstruction artifacts~\cite{Eckert_2018}.
Each high-angle illumination in aIDT is very sensitive to high and low-frequency object information, so calibrating the angle is critical to recover volumetric information without error.
The algorithm imposed here follows two geometric constraints.
First, the distribution of our LED ring is expected to obey a circular geometry
\begin{equation}\label{Eq3}
u_i^2 + v_i^2 = (\mathrm{NA}_{\mathrm{obj}}/\lambda)^2,
\end{equation}
where $\left( {{u_i},{v_i}} \right)$ denote the estimated spatial frequency for the $i$th LED.
Second, the LEDs are expected to be equally spaced.
Correspondingly, each pair of neighboring LEDs occupy a $\pi/12$ radian angular space.
The initial estimate of the angular coordinates $\theta_i$ of each LED follows
\begin{equation}\label{Eq4}
{\theta _i} = \mathrm{atan2}(v_i/u_i), \text{and}~{\theta _i} \in \left[ {-\pi:\frac{{\pi }}{{12}}:\pi } \right),
\end{equation}
where $\mathrm{atan2}$ computes the angle of the inverse tangent function in the unit of radian.

Our self-calibration algorithm starts with an initial guess $u_i^{\mathrm{init}}, v_i^{\mathrm{init}}$ from the algorithm in \cite{Eckert_2018}, whose estimated LED positions are often contaminated by noise.
Accordingly, the final calibrated LED positions $u_i^{\mathrm{cal}}, v_i^{\mathrm{cal}}$ are parameterized as
\begin{equation}\label{Eq5}
\begin{aligned}
& u_i^{\mathrm{cal}} = \Delta u + \mathrm{NA}_{\mathrm{obj}}\cos \left( {{\theta _i} + \Delta \theta }\right)/\lambda, \\
& v_i^{\mathrm{cal}} = \Delta v + \mathrm{NA}_{\mathrm{obj}}\sin \left( {{\theta _i} + \Delta \theta } \right)
/\lambda,
\end{aligned}
\end{equation}
and are found by solving the optimization problem
\begin{equation}
\min_{u_i^{\mathrm{cal}}, v_i^{\mathrm{cal}}}
\sum\limits_{i = 1}^{24}
\bigg[{{\left( u_i^{\mathrm{cal}} - u_i^{\mathrm{init}} \right)}^2}
+ {\left( v_i^{\mathrm{cal}}- v_i^{\mathrm{init}} \right)^2}\bigg].\label{Eq6}
\end{equation}



\section{Results}

\subsection{Angle self-calibration and performance characterization}
\begin{figure*}[!ht]
    \centering
    \includegraphics[width=0.8\textwidth]{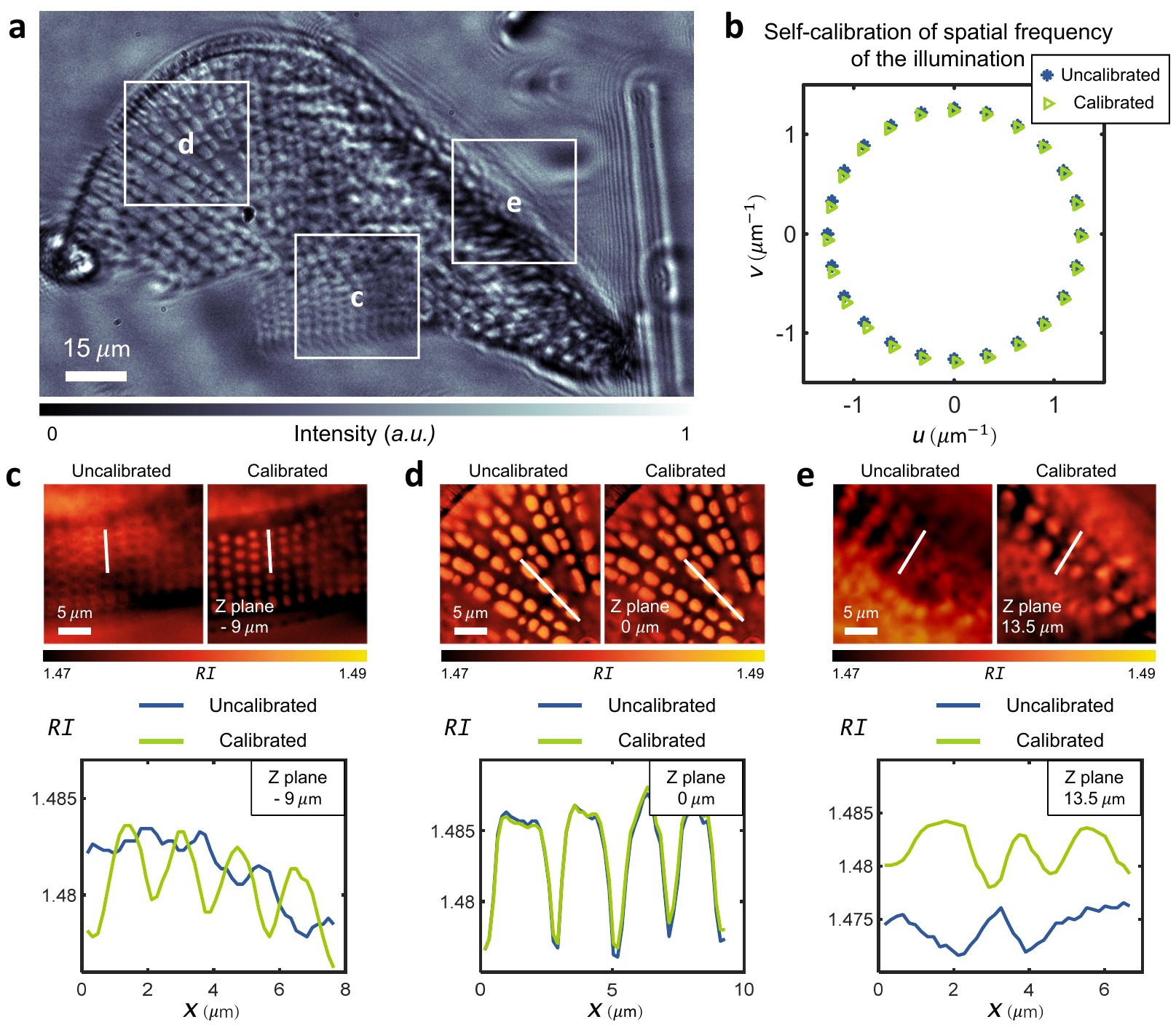}
    \caption{Results from our LED position self-calibration method.
     (a) A sample intensity image of a diatom under a certain single-LED illumination.
     (b) LED positions from manual alignment (termed uncalibrated, marked in blue star) and our self-calibration methods (termed calibrated, marked in green triangle) as plotted in the spatial frequency coordinates.
     (c-e) Comparison of the reconstructed RI slides before and after calibration.
     (c) $z=-9\mu$m, (d) $z=0\mu$m, and (e) $z=13.5\mu$m.
     More detailed comparisons across the whole volume is provided in Video 2
     (Video 2, MPEG, 2.6 MB).
     }
    \label{Fig2}
\end{figure*}

Achieving high-quality RI tomographic reconstruction requires accurate LED positioning, especially when imaging large-volume objects under high NA illuminations.
In practice, removing all residual errors in the LED positions using only manual alignment and physical calibration procedures is non-trivial.
We instead develop an algorithmic self-calibration method for finely tuning the LED positions and demonstrate our technique's improvement of 3D reconstructions.

Our self-calibration method combines two main principles for high-accuracy measurements.
First, our TF analysis shows that each intensity image's Fourier spectrum should contain distinct circular regions with center positions defining the illumination angle.
A demonstration of this principle on experimental data is shown in Video 1.
A previously developed algorithm~\cite{Eckert_2018} already utilizes these features, so we adopted this technique to provide initial LED position estimates.
In practice, this algorithm's susceptibility to noise can introduce position error exceeding the illumination unit's engineering and alignment tolerance.
To correct this error, we subsequently incorporate two geometric constraints refining the LED positions to form a ring shape with equal angular spacing using a nonlinear fitting algorithm.
This is warranted because the surface-mount technology has high position accuracy of device placement in printed circuit board manufacture process and our illumination unit's printed LED circuits have expected engineering tolerances below 5$\mu$m generally.

We demonstrate the effectiveness of this method on diatom microalgae (S68786, Fisher Scientific) fixed in glycerine gelatin imaged with a 0.65 NA MO.
An example intensity image is shown in Fig.~\ref{Fig2}a.
The low-absorbing features (i.e. ``phase'' features) are already visible due to asymmetric illumination, akin to differential phase contrast~\cite{Mehta2009}.
Figure~\ref{Fig2}b compares the LED positions from manual alignment (blue stars) and our self-calibration method (green triangles) (More details are illustrated in Fig. S2 in supplementary material).
The RI reconstruction improvements from our technique are shown in three outsets highlighting features located at different axial positions.

As shown in Fig.~\ref{Fig2}d, the LED mis-calibrations have minimal effect for structure reconstructions at the objective's focal plane ($z=0\mu$m).
Significant RI degradation from incorrect illumination angles is observed at defocus reconstruction planes (Fig.~\ref{Fig2}c,e).
This degradation is intuitively explained under the ``light field''~\cite{Tian.Waller2015,Tian2014a} effect: for a fixed angular error, a larger defocus induces a larger feature displacement error.
Our self-calibration method largely mitigates these errors to provide high-quality RI reconstructions (Fig.~\ref{Fig2}c-e).
Both the lateral resolution and contrast are preserved across the entire volume and recovers diatom frustules previously lost at large defocus due to mis-calibration (Fig.~\ref{Fig2}e).
This calibration procedure allows us to provide high-quality RI distributions with aIDT and is used in our subsequent results.

\subsection{Tomographic characterization of Surirella spiralis}

\begin{figure*}[!t]
    \centering
    \includegraphics[width=0.87\textwidth]{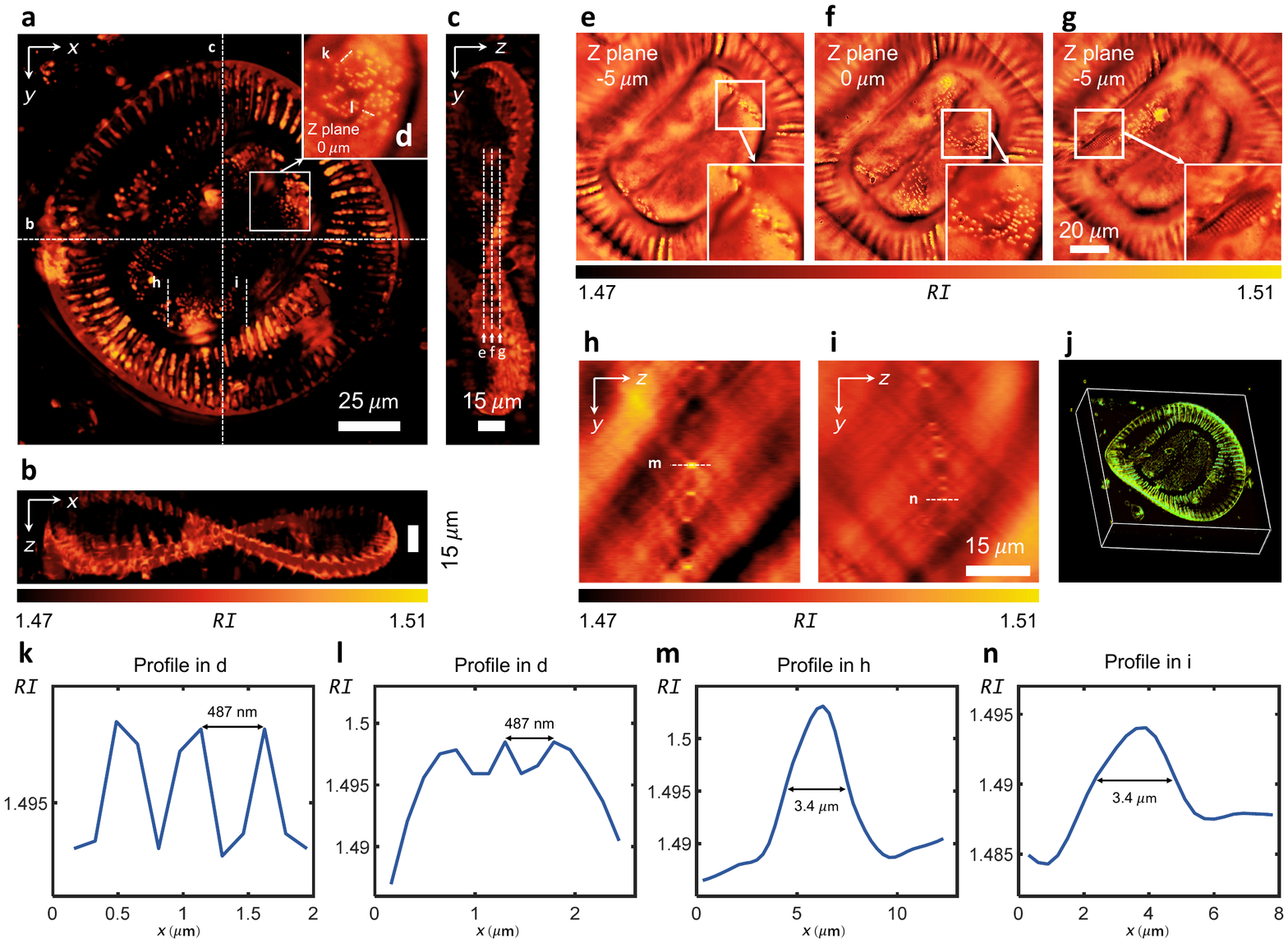}
    \caption{RI tomography of Surirella spiralis.
    (a-c) The maximum RI projection views of the recovered 3D RI distribution in the $x-y$, $x-z$, and $y-z$ planes.
    (d) Zoom-in on closely packed frustule structures.
    (e-g) Reconstructed 2D cross sectional RI slices at -5$\mu$m, 0$\mu$m and 5$\mu$m planes.
    (h-i) YZ-cross sectional views of the reconstructed RI.
    (j) A 3D rendering view of the reconstructed RI distribution.
    Additional cross-sectional reconstruction and 3D rendering view from different perspectives are shown in Video 3.
    (k-n) Line profiles across frustule structures to quantify the reconstructed lateral and axial resolution
    (Video 3, MPEG, 3.3 MB). }
    \label{Fig3}
\end{figure*}

We demonstrate aIDT's ability to characterize complex single cell organisms with intracellular resolution on a Surirella spiralis diatom sample.
We acquired 24 intensity images under oblique illuminations and reconstructed the sample's RI across a 50 $\mu$m volume as shown in Fig.~\ref{Fig3}.
The benefit of this technique is clearly shown in its recovery of multi-scale features of the sample.
aIDT recovers the characteristic Surirella spiralis saddle shape spanning the full 50 $\mu$m reconstructed volume (Fig.~\ref{Fig3}a-c).
Within this large saddle, fine structures including silica frustules are also visible and well-resolved across multiple reconstructed slices (Fig.~\ref{Fig3}d-g) and the YZ cross-sectional views (Fig.~\ref{Fig3}h-i).
To further illustrate this structure, Fig.~\ref{Fig3}(j) displays the RI rendered as a 3D volume \cite{schindelin2012fiji} of Surirella spiralis. Line profiles across these 10 $\mu$m tall frustules in Fig.~\ref{Fig3}(k-n) demonstrate near diffraction-limited lateral resolution of 487 nm and axial resolution of 3.4 $\mu$m.

aIDT quantitatively recovers both full cell-sized features and intracellular structures easily using a small set of intensity images from a single focal plane.
These results make aIDT advantageous for biological research applications
containing complex environments requiring simultaneous, multi-scale sample evaluation.
In addition, the lack of sample scanning with this technique increases its utility for dynamic sample imaging where living objects easily move out-of-focus.
We show aIDT's application to both of these cases in the subsequent sections.

\subsection{RI tomography on cell clusters}

\begin{figure}[!t]
    \centering
    \includegraphics[width=0.48\textwidth]{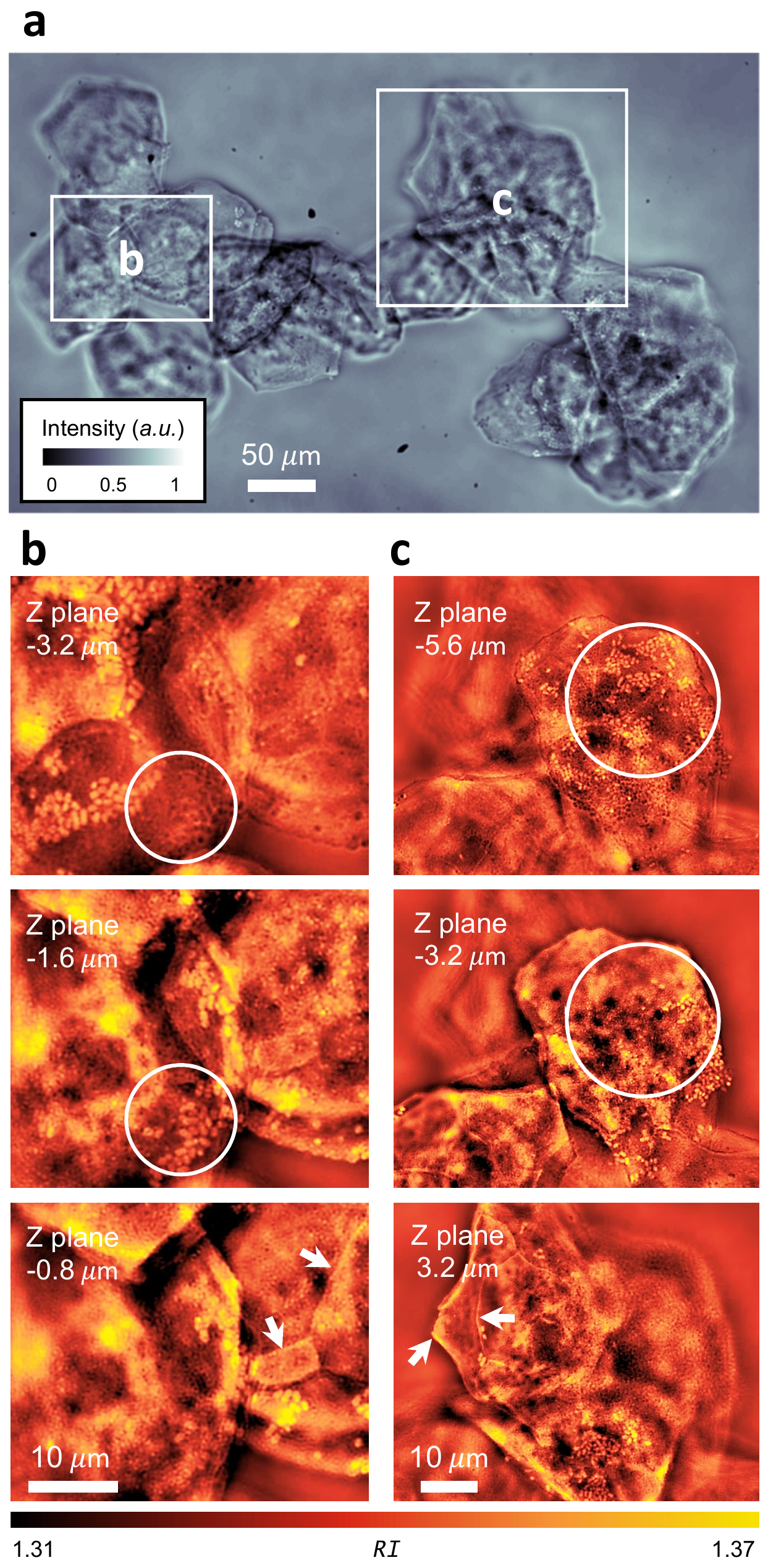}
    \caption{Single cell RI tomography of  unstained human cheek cell clusters.
    (a) A sample raw intensity image under  annular illumination.
    (b,c) Reconstructed RI cross-sections demonstrate the sectioning capability enabled by the aIDT.
     Additional examples are shown in Video 4
     (Video 4, MPEG, 2.8 MB).
    }
    \label{Fig4}
\end{figure}

We next apply aIDT to evaluating complex biological cell clusters and environments.
Existing 2D phase imaging techniques are often used when imaging monolayers of cells.
These integrated phase map techniques contain less useful information, however, when imaging cell clusters more commonly found in biological systems.
Our aIDT technique overcomes this problem by recovering multiple, independent RI cross-sections across extended volumes.
This approach enables better depth-sectioning of the sample such that larger biological structures with greater complexity can be evaluated without significant information loss.

We demonstrate this ability to recover complex biological environments on clusters of unstained human epithelial buccal cells distributed on a glass sample slide.
A sample normalized intensity image is shown in Fig.~\ref{Fig4}a showing the cell cluster's complexity and its defocused regions highlighting the sample's large volume.
We take 24 intensity images and reconstruct the RI across a 16 $\mu$m volume.
We expand two regions of the reconstructed RI in Fig.~\ref{Fig4}b,c highlighting our depth-sectioned reconstructions.

The benefit of aIDT when imaging complex environments is seen in its high-resolution reconstructions across the entire cell volume.
At each reconstructed slice we observe cellular membrane folds, cell boundaries, nuclei, and intracellular features with high resolution (Fig.~\ref{Fig4}b,c arrows).
In addition, we recover native bacteria, likely a staphylococcus strain, distributed on the cells throughout the sample (Fig.~\ref{Fig4}b,c circles).

Quantifying the 3D RI distribution of entire cells, their subcellular structures, and external environment features such as bacteria has significant potential in biological research applications.
The recovered volumetric RI distributions of cellular features enables the calculation of dry and buoyant mass, sphericity, and other morphometric descriptors used for cell profiling \cite{lobo2016insight, zangle2014live}.
Because subcellular and bacterial structures are also resolved, these parameters can be applied to subcellular features with aIDT.
With aIDT's fast acquisition rates and large volume recovery, shown experimentally in the next section, longitudinal maps of structure mass and volume changes can be mapped in real-time throughout multi-cellular complex environments.
Quantifying these factors could be highly beneficial to immunology and pathophysiology applications ,where longitudinal studies of parasite and bacterial interactions and induced morphological changes in cells carry critical information for understanding and mitigating infection \cite{chandramohanadas2011biophysics,sadrearhami2019antibiofilm}.
Furthermore, quantifying volumetric morphological changes of cellular and subcellular information also has significant utility in oncology for both differentiating cancer types and evaluating their response to drug and therapy treatments \cite{sung2012stain,li2019quantitative,zhikhoreva2018morphological}.

\subsection{Dynamic RI tomography of C. elegans in vitro}

\begin{figure*}[!h]
    \centering
    \includegraphics[width=0.9\textwidth]{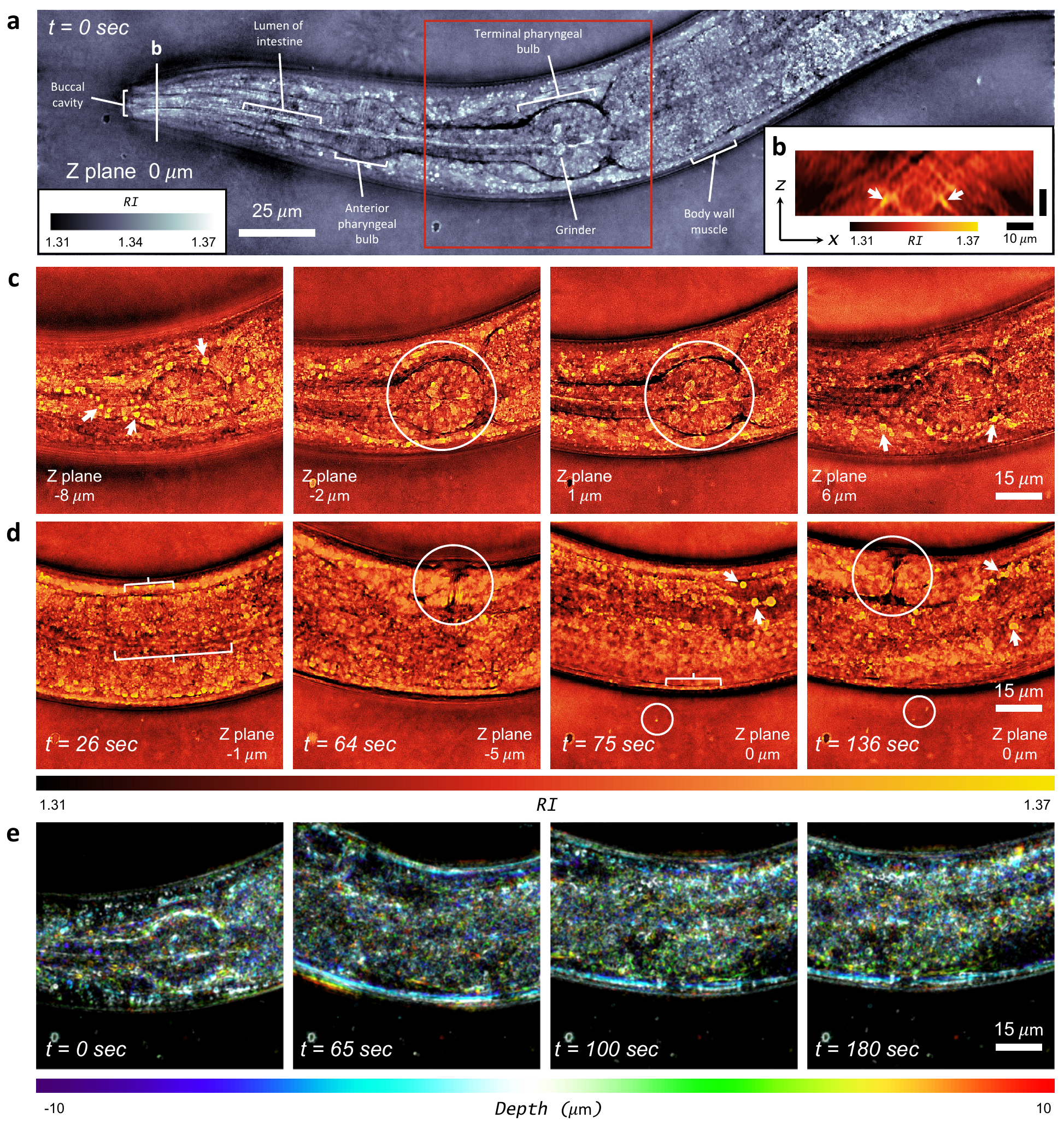}
    \caption{Time-lapse \emph{in vitro} tomographic imaging of \emph{C. elegans}.
    (a) Recovered RI slice located at central plane at \emph{t = 0 sec}. The full \emph{C. elegans} worm reconstruction visualization is shown in Video 5.
    (b) RI stack section in $x-z$ plane close to the mouth of \emph{C. elegans}. Buccal cavity of \emph{C. elegans} is distinguishable (indicated by the white arrows).
    (c) RI distribution of worm at different $z$ planes in the marked red square region at t = 0 sec. Time-lapse details are demonstrated in Video 6.
    (d) Visualizations and RI quantification of the \emph{C. elegans} internal tissue structures at different time points and axial planes.
    (e) Depth color coding of 3D RI measurements of sample in the selected sub-region with fix position in the field of view. 4 different time points to illustrate the time lapse results of \emph{C. elegans}
    (Video 5, MPEG, 9.7 MB; Video 6, MPEG, 9.5 MB).
    }
    \label{Fig5}
\end{figure*}

A major advancement enabled by aIDT is the ability to perform high-speed \emph{in vitro} tomographic imaging of biological samples using a small number of intensity-only measurements.
This allows us to visualize 3D dynamical biological phenomena with minimal motion artifacts, which is particularly challenging using existing RI tomography techniques.
We demonstrate this ability on unstained, live \emph{C. elegans} worms \cite{stiernagle2006maintenance,kim2013long} at a 10.6 Hz volume rate.
We image a volume containing 333$\times$98$\times$21 $\mu$m$^3$.
In a time-lapse series, each image stack includes 8 frames (for reconstructing each RI volume) that were recorded with a 4.4 ms exposure time over a 3 min period to evaluate fast motions in a living \emph{C. elegans}.

The reconstructed RI of the \emph{C. elegans} worm is shown in Fig.~\ref{Fig5}.
Reconstructed RI $x-y$ and $x-z$ cross-sections at the $z=0\mu$m plane at $t = 0$ sec are shown in Fig.~\ref{Fig5}a and Fig.~\ref{Fig5}b, respectively.
Figure~\ref{Fig5}c shows the RI distribution of the worm at different $z$ planes in the marked region at $t = 0$ sec.
Figure~\ref{Fig5}d illustrates the RI distribution of the \emph{C. elegans} internal tissue structures at different time points and axial planes.
Depth-coded projections of our reconstructions are also provided in Fig.~\ref{Fig5}e, where the volumetric RI distribution is shown for several different time points.
The full \emph{C. elegans} worm reconstruction visualization is shown in Video 5.
The results show that aIDT is robust to motion artifacts and resolves internal features during high-speed worm motion, as clearly demonstrated in Video 6.

Our technique easily visualizes and provides RI quantification of the \emph{C. elegans} internal tissues.
The anterior and terminal pharyngeal bulbs are clearly resolved in our reconstruction (Fig.~\ref{Fig5}a,c) as well as the grinder and intestines (Fig.~\ref{Fig5}c circles, Fig.~\ref{Fig5}d long bracket).
Lipid droplets and lysosomes are also distinguished in the worm head at different axial layers (Fig.~\ref{Fig5}c arrows).
Within the worm body, we recover the vulva (Fig.~\ref{Fig5}d circles) across multiple axial slices, body wall muscles (Fig.~\ref{Fig5}d short brackets), and features resembling the worm's nerve cord (Fig.~\ref{Fig5}d).
We also observe \emph{E. coli} bacteria living and moving independently of the \emph{C. elegans} (Fig.~\ref{Fig5}d small circles).
Additional results on fixed \emph{C. elegans} can be found in supplementary material and Video S2.

aIDT enables a simple, label-free approach for volumetric imaging in the biological research community.
The tissues shown in Fig.~\ref{Fig5} and Video 5 often undergo phenotypic changes from genetic mutations during biological studies~\cite{corsi2015transparent}.
Quantifying these tissue changes and studying their effect on live worm behavior in a natural, label-free setting would be highly beneficial in understanding the effects of targeted genetic mutations on living organisms.
Because our technique captures bacteria motion concurrently with the \emph{C. elegans}, aIDT could also evaluate multi-organism interactions and provide 3D bacteria tracking during longitudinal studies.
The versatility of this technique in visualizing multiple tissue types means it has utility spanning from neurology to pathogenesis and wound healing~\cite{corsi2015transparent}.

\subsection{aIDT reconstruction accuracy and sensitivity analysis}
\begin{figure*}[!h]
    \centering
    \includegraphics[width=0.75\textwidth]{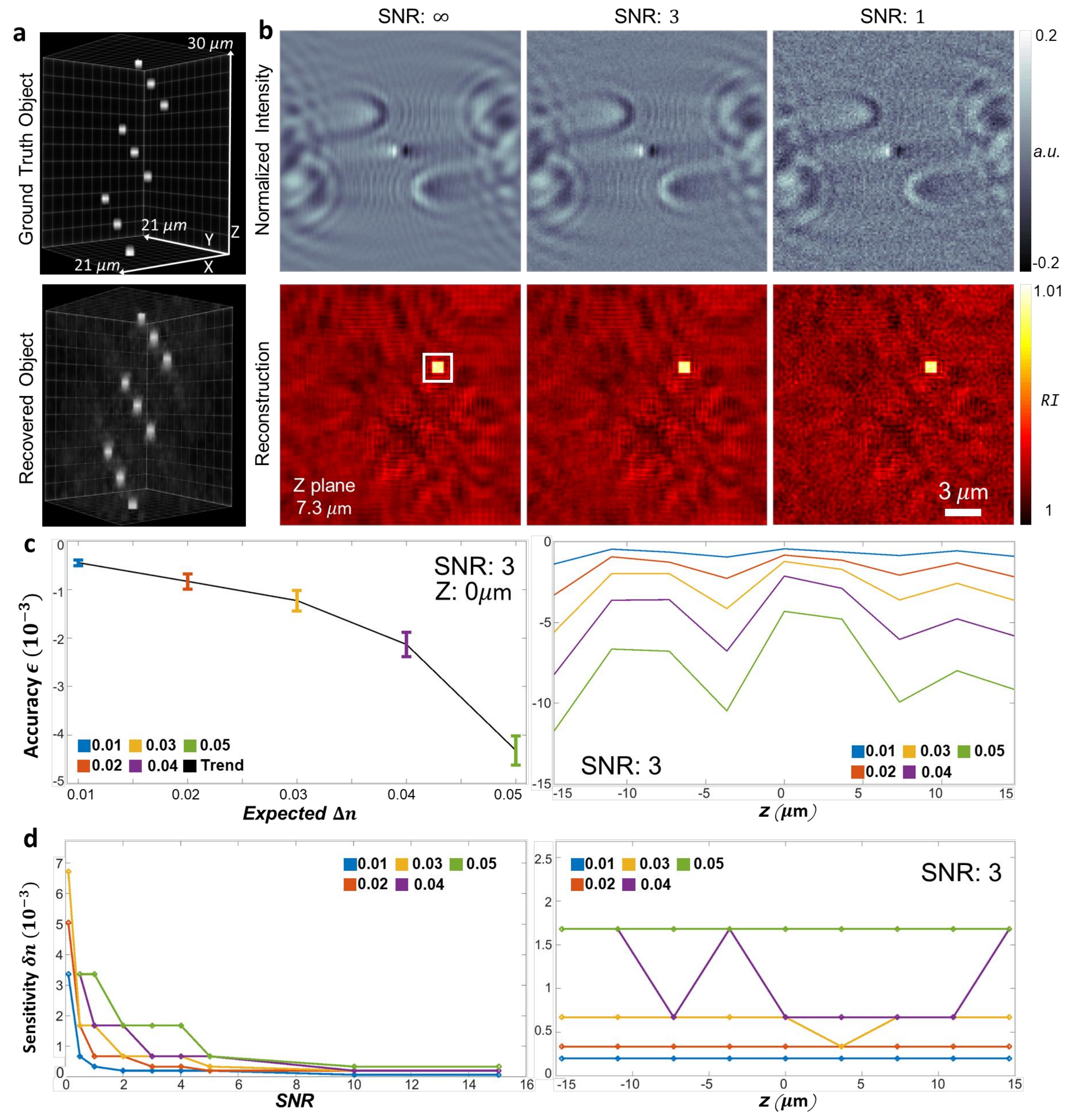}
    \caption{Simulation for quantifying aIDT accuracy and sensitivity .
    (a) Top: The object consists of a cuboid array occupying a $21\mu$m$\times21\mu$m$\times30\mu$m volume.   Bottom: an example aIDT reconstruction.
    (b) Top: Simulated intensity images with decreasing SNR.  Bottom: aIDT reconstructions at Z$=7.3 \mu$m. The white box indicates the region over which the cuboid's RI is obtained for accuracy and sensitivity analysis.
    (c) Left: aIDT accuracy ($\epsilon$) evaluations across the object RI. The plot shows the average difference between the aIDT reconstruction and true RI across 100 realizations. Error bars show the standard deviation of this difference. Right: aIDT accuracy across the simulated axial range. aIDT provides accurate RI recovery under low contrast ($\Delta n =0.01-0.03$) objects and loses accuracy from highly scattering features ($\Delta n =0.05$).
    The accuracy is stable across the entire reconstruction volume but exhibits fluctuations due to boundary effects in the computation.
    (d) aIDT sensitivity ($\delta n$)  analysis as a function of SNR (Left) and axial position (Right) under the experimental SNR.
    aIDT's sensitivity to small RI changes varies with the object's RI contrast but maintains  to be better than $\delta n = 0.002$ even at high RI contrast for SNR = 3.
    The sensitivity is constant along $z$ for low-contrast objects but fluctuates with increasing RI contrast.}
    \label{Fig6}
\end{figure*}

Having shown aIDT's utility for label-free dynamic biological sample imaging, we further evaluate the modality's {\it accuracy} and {\it sensitivity}.
While the experimentally recovered volumes exhibit RI ranges matching expected biological values, the inherent variability of these specimens prevents quantitative analysis of the system's accuracy and sensitivity for recovering the true RI distribution and detecting small RI variations, respectively.
These parameters were briefly explored for conventional IDT~\cite{ling2018high}, but the lack of manufactured, well-characterized objects limited the accuracy and sensitivity analysis to thin glass structures with high-contrast RI distributions.
These structures are not representative of most biological samples' RI range or size, and their high-contrast nature generates multiple-scattering behavior that invalidates the IDT model.
Recent works~\cite{hosseini2016,juffmann2020local} show such experimental sensitivity analyses are possible in quantitative phase systems with rigorous testing using expensive hardware that was not readily available for the aIDT system.
Thus, determining aIDT's accuracy and sensitivity is a challenging task.
Here, we instead evaluate aIDT in simulation to determine its theoretical accuracy and sensitivity over the RI range present in our experimental data.

Our simulations were performed in MATLAB with three primary components: 1) a ground-truth object, 2) a rigorous forward model simulating the field through the object, and 3) our aIDT inversion algorithm.
For the object, we generated 3$\times$3 cuboid arrays inside a 21$\times$21$\times$30$\mu$m$^3$ volume with variable RI (Fig.~\ref{Fig6}a).
Each cuboid occupied a 0.97$\times$ 0.97$\times$1.2$\mu$m$^3$ volume and was spatially separated by $3.25 \mu$m and $2.4 \mu$m in lateral and axial dimensions, respectively.
This separation recovers a single cuboid for each reconstructed aIDT slice over the same volume considered in our experiment.
For the cuboid RI, we assumed a homogeneous imaging medium ($n_m=1$) and generated arrays with RI range $RI=[1.0033, 1.0567]$ following the equation $RI = n_{\mathrm{base}} + \delta n$, where $n_{\mathrm{base}} = [1.01,1.02,1.03,1.04,1.05]$ and $\delta n = [-.0067, 0.0067]$.
This large RI range allowed us to evaluate aIDT's accuracy over the contrast range observed in our experiments ($\Delta RI \leq 0.05$) and test aIDT's sensitivity to small RI variations at each $n_{\mathrm{base}}$ level.
Here, the selected range of $\delta n$ values corresponded to object phase variations between 1 and 100 mrad following $\phi = 2\pi\lambda^{-1}\Delta n\Delta h $.
These parameters allowed the evaluation of both aIDT's accuracy and its sensitivity to small RI changes across a large contrast range.

With these objects, we simulated aIDT intensity images using the convergent Born series model~\cite{Osnabrugge2016}.
This forward model efficiently simulates multiple-scattering through large object volumes using a convergent Born series expansion, making it ideal for evaluating aIDT's recovery capabilities.
Using the illumination angles from our 8-LED illumination aIDT case, we simulated the scattered field through the cuboid array and propagated the final field through a 0.65 NA, 40$\times$ objective lens to obtain our intensity image stack.
We repeated this simulation process for each cuboid array with differing refractive index and reconstruct the object volume using our aIDT algorithm.
Furthermore, we added white Gaussian noise to the intensity images generating a Signal-to-Noise Ratio (SNR) ranging from 0 to 15, and generate 100 realizations for each SNR level.
The SNR is quantified by the ratio between the signal contrast and noise level as SNR$ = \sigma_{\mathrm{Signal}}/\sigma_{\mathrm{Noise}}$, where $\sigma$ denotes the standard deviation.
To determine the reconstruction accuracy, we compared the median recovered RI over each cuboid area (Fig.~\ref{Fig6}b, white square) with the ground-truth (GT) object filtered to match the reconstruction bandwidth.
For the reconstruction sensitivity, we evaluated the separation between the small RI variations $\delta n$ from the central RI value $n_{\mathrm{base}}$.
We used the Ashman's D test~\cite{Ashman1994} for separating bimodal distributions to determine the minimum RI variation detectable for each SNR condition and considered two RI values to be separable when $D > 2$.

The simulation results for accuracy and sensitivity are summarized in Fig.~\ref{Fig6}c and d, respectively.
Under SNR matching our experimental condition, Fig.~\ref{Fig6}c shows the average RI mismatch between our reconstruction and the ground truth across RI (Left) and axial position (Right) over the 100 realizations simulated for this SNR condition.
The error bars show the standard deviation in this mismatch over these realizations.
We obtain nearly equivalent RI recovery under low contrast ($\Delta n = 0.01-0.03$) and large underestimations for high-contrast objects at $\Delta n=0.05$.
Underestimations for large RI contrast objects was expected due to the presence of multiple-scattering invalidating the assumption of weak scattering in the aIDT model.
An offset still exists between aIDT and the GT object for weakly scattering objects, which indicates the approximations we make in the aIDT model do reduce our system's accuracy.
Across different axial positions, we observe mostly constant $\epsilon$ with greater offsets occurring for high-contrast features.
Across all RI cases, we observe a periodic loss in accuracy over the tested axial range.
This periodic loss corresponds to the cuboid appearing close to the volume boundary, suggesting that boundary condition issues exist in our simulation.
Under weakly scattering conditions, we observe that this axial-dependent behavior is within our model's
$\epsilon$ offset and are considered to be minimal.
We will solve these boundary condition issues in future work.

Figure~\ref{Fig6}d shows aIDT's theoretical sensitivity range over the imaging volume and across multiple SNR conditions.
At SNR = 3, we have a minimum sensitivity to $\delta n=2\times10^{-4}$ for low-contrast objects and $\delta n = 2\times10^{-3}$ for high-contrast features.
Across different axial positions, aIDT's sensitivity shows mostly static sensitivity with greater fluctuations for objects with stronger contrast.
These results indicate that aIDT exhibits high sensitivity to RI variations across the full reconstructed volume under low-contrast imaging conditions.

Our simulations show aIDT can provide high-accuracy and high-sensitivity RI recovery of volumetric biological samples under the proper conditions.
Given weakly scattering samples within our model's validity range, aIDT can recover correct accurate RI values and detect small fluctuations to variations in the object's RI.
This analysis is promising for biological sample evaluation where these small RI variations could correspond to the presence of pathogens in cells~\cite{chandramohanadas2011biophysics}.
While this accuracy and sensitivity will suffer from experimental factors including objective aberrations and illumination misalignments, our simulations shown here indicate aIDT provides accurate, highly-sensitive volumetric recoveries of biological samples.

\section{Discussion}
We introduced aIDT, a high-speed, label-free, scanless non-interferometry based quantitative imaging tool for the 3D evaluation of unlabeled weakly scattering specimens.
By combining an LED ring illumination unit with a standard brightfield microscope, we capture obliquely illuminated intensity images and perform 3D deconvolution to recover the slice-wise 3D RI distribution.
The geometry fitting between illumination angle with the objective NA optimally encodes both low and high spatial frequencies into each acquired image.
This illumination scheme reduces the system's data requirement and allows us to image large 3D volumes of weakly-scattering samples at high speeds.
We demonstrated the success of aIDT on various biological samples, from fixed microalgae, cheek cells, to living \emph{C. elegans}.
Finally, we showed aIDT's has high theoretical accuracy and sensitivity limits in simulation under a range of noisy imaging conditions.
We believe this method will set an excellent foundation for other research projects and applications, and the aIDT has the potential as a tool of great biological interest by showing its use in monitoring cell morphology and dynamics in noninvasive high-speed measurements.

Due to the full usage of objective NA, the achievable phase imaging resolution can be extended to the incoherent diffraction limit.
And the proposed technique is mainly focus on high-speed $\emph{in vitro}$ biological sample imaging, so the boundary of RI mapping resolution of this work is the tradeoff between the working distance and NA of the objective.
The quantification of phase sensitivity is important for aIDT imaging system, and angle calibration quality, object RI, and the assumed slice thickness will effect the aIDT’s sensitivity.
 But it requires more complicated setups and control samples to experimentally evaluate the system sensitivity, and more detailed analysis and enchantment of phase sensitivity are beyond the scope of this work.

Our IDT model is currently limited by the single scattering approximation that ignores information encoded in the multiple scattering.
Recently, several groups have demonstrated  multiple scattering models suitable for solving large-scale imaging problems~\cite{Tian.Waller2015,kamilov2015learning,Chowdhury_2019,Lim_2019}, which will be considered in our future work.
Our model-based reconstruction approach is also constrained by unknown experimental variabilities that are difficult to be fully parameterized via an analytical model, which may be overcome using emerging learning-based inversion techniques~\cite{nguyen2018deep,xue2019reliable,rivenson2018phase,sinha2017lensless,Li_2018,Feng_2019,Lyu_2019,Barbastathis_2019}.

\section*{Acknowledgments}
We thank Dr. Christopher Gabel, Dr. Daniel Taub, Dr. Gregory Wirak for providing biological samples.

\section*{Supplemental Documents}
We provide example datasets and an open source implementation of aIDT at \href{https://github.com/bu-cisl/IDT-using-Annular-Illumination}{https://github.com/bu-cisl/IDT-using-Annular-Illumination} and see supplementary material for supporting content.

\end{document}